# Giant electrostrain of 0.57% in a periodically orthogonal poled lead titanate zirconate ceramic via reversible domain switching

Faxin Li[1,2,a], Qiangzhong Wang[1], Hongchen Miao[1]

[1]LTCS and College of Engineering, Peking University, Beijing, 100871, China

[2]Center for Applied Physics and Technology, Peking University, Beijing, China

**Abstract**

The widely used ferroelectric ceramics based actuators always suffer from small output strains (typically ~0.1-0.15%). Non-180° domain switching can generate large strain in ferroelectrics but it is usually irreversible. In this work, we tailored the domain structures in a soft lead titanate zirconate (PZT) ceramic by periodically orthogonal poling. The non-180° switching in this domain-engineered PZT ceramics turns to be reversible, resulting in giant electrostrains up to 0.57% under a field of 2kV/mm (dynamic $d_{33}^*$(=S/E) of 2850pm/V). The large electrostrain keeps quite stable and even slightly increases after $10^4$ cycles of loading, which is very promising for next-generation large-strain actuators.

[a]Author to whom all correspondence should be addressed, Email: lifaxin@pku.edu.cn

# 1. Introduction

Ferroelectric/piezoelectric materials have been widely used in actuation areas for their quick responses, compact size and large blocking force [1]. The currently used lead titanate zirconate (PZT) based stack actuator is always suffering from its small output strain(typically ~0.1-0.15%), which is much smaller than that by the shape memory alloys (~several percent) [2] and also smaller than that by the giant magnetostrictive materials, Terfenol-D (~0.2%) [3]. In the past decades, various approaches have been adopted to enhance the electrostrains in ferroelectrics which can mainly be classified into three types: i) Developing relaxor ferroelectric single crystals [4-8]. The electrostrains can be enhanced to be over 1% in these crystals via electric field induced ferroelectric phase transitions. However, the high cost of single crystals and the quick drop of electrostrain upon applying a moderate prestress[6-8] make these materials not suitable for industrial actuations. ii) Developing lead-free piezoelectric ceramics [9-12]. The electrostrains in these materials can reach up to 0.7% with very large hysteresis under a large field of 5kV/mm [11]. While under the typical actuation field of 2kV/mm, currently the strain in lead-free ceramics can at most reach 0.2% [12], showing no obvious advantage compared to PZT. iii) Making non-180° domain switching reversible via point defect mediating [13,14] or electromechanical loading [15-17]. The large reversible strain in this way can reach up to 0.93% in BaTiO3 crystals and 0.66% in $Pb(Mn_{1/3}Nb_{2/3})O_3$-$PbTiO_3$ crystals [17]. However, after cycles of operations, the electrostrains either greatly decreases due to the relaxation of the internal bias field [13,14] or the crystals turns to be broken because of the large generated internal stresses [17]. In this work, we show that large reversible switching strain can also be achieved in ferroelectric ceramics. By generating engineered domain structures in a PZT-5H ceramics via periodically orthogonal poling, the non-180° domain switching turns to be reversible due to the self-generated interfacial stresses between the adjacent regions with different poling directions. The reversible switching strain can reach up to 0.57% under a field of 2kV/mm and is quite stable after cycles of operations, which is very promising for next-generation large-strain actuators.

## 2. Experimental

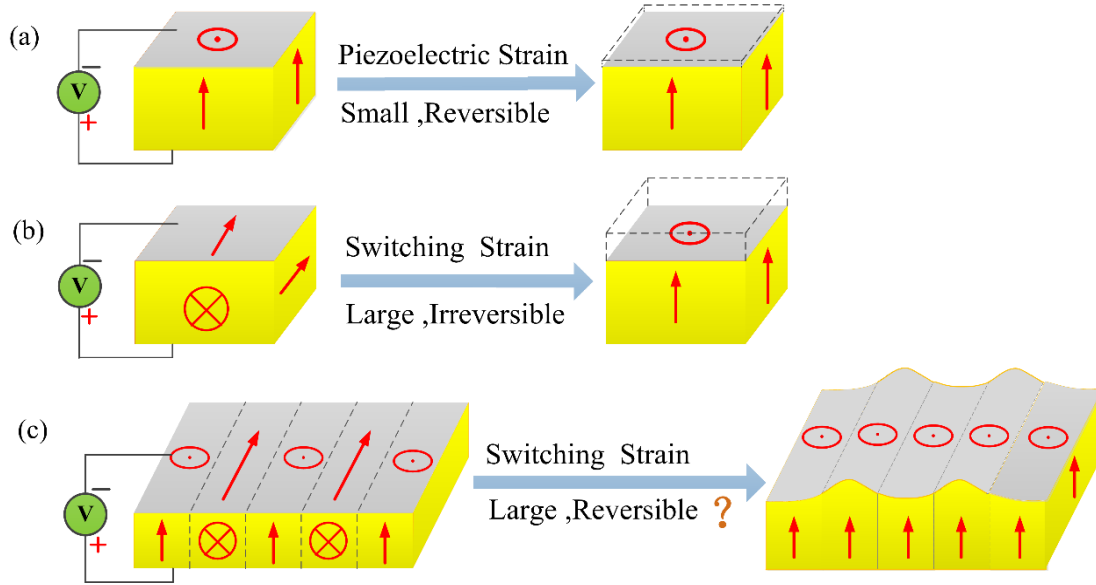

Fig.1 Principle of possibly generating large reversible strain in ferroelectric ceramics. (a) Small and reversible piezoelectric strain in conventional piezoelectric actuators; (b) Large but irreversible strain in ferroelectrics via non-180°domain switching; 3) Large and reversible (?) switching strain in a periodically orthogonal poled ferroelectric ceramics.

The principle of possibly generating large reversible strains in ferroelectric ceramics is illustrated in Fig.1. When an electric field is applied on a poled ferroelectric ceramics along the poling direction, a small reversible strain can be achieved via the piezoelectric effect, as shown in Fig.1(a). If a large electric field perpendicular to the original poling direction is applied, as shown in Fig.1(b), a large but irreversible strain will be generated due to the non-180°domain switching. Then how about applying a large electric field on a periodically orthogonal poled ferroelectric ceramics as shown in Fig.1(c)? For convenience, the vertical poled region in Fig.1(c) is denoted as Region C and the in-plane poled region denoted as Region A. Obviously, a fairly large switching strain can be obtained in Region A under a large electric field, however, can this switching strain be reversible? To investigate this, we fabricated the periodically poled ferroelectric ceramics and conducted electrostrain measurements.

A soft PZT-5H ceramics is used as the raw material and its properties can be found elsewhere [18]. The sample was cut into blocks with the dimensions of 10mm × 4mm × 2mm and firstly poled along the thickness (2mm) direction. Then the sample was evenly divided into five regions along the length direction (10mm) with the period of 2mm, and the two even-number regions was in-plane poled along the width direction (4mm), as shown in Fig.1(c). Finally, the lateral electrodes were removed and the top/bottom electrodes were re-spread for electric loading.

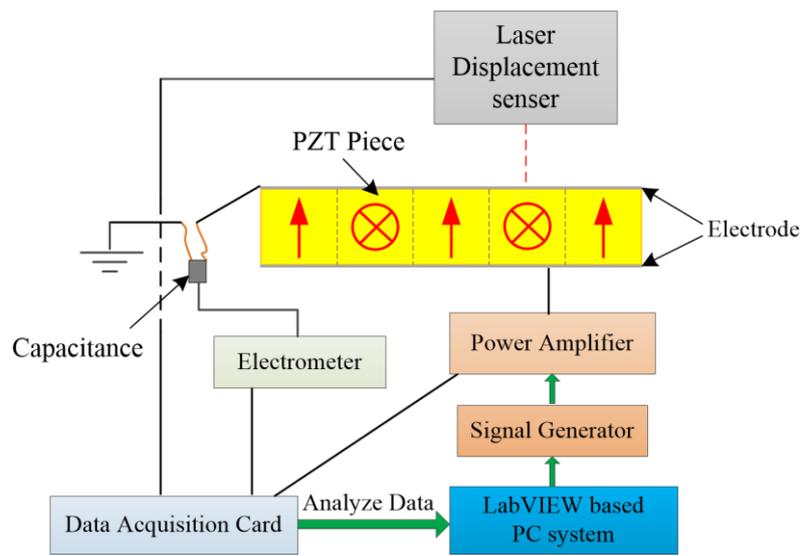

Fig.2 Testing setup for measuring the electrostrains.

The electrostrain testing setup was shown in Fig.2. A uni-polar voltage signal generated by a Functional Generator (Agilent 33220A) and amplifier by a high-voltage amplifier (Trek MODEL 609B) was used to apply the large electric field on the periodically poled PZT sample. A laser displacement sensor (LK-G30) with the resolution of 0.01um was used to measure the displacement and an electrometer (EST 103) to monitor the charge variations during electric loading. The signals of displacement and charge were collected by using a Data Acquisition Card and then converted to strain and polarization. All the testing was controlled and monitored by a Labview-based PC system.

## 3. Results and Discussions

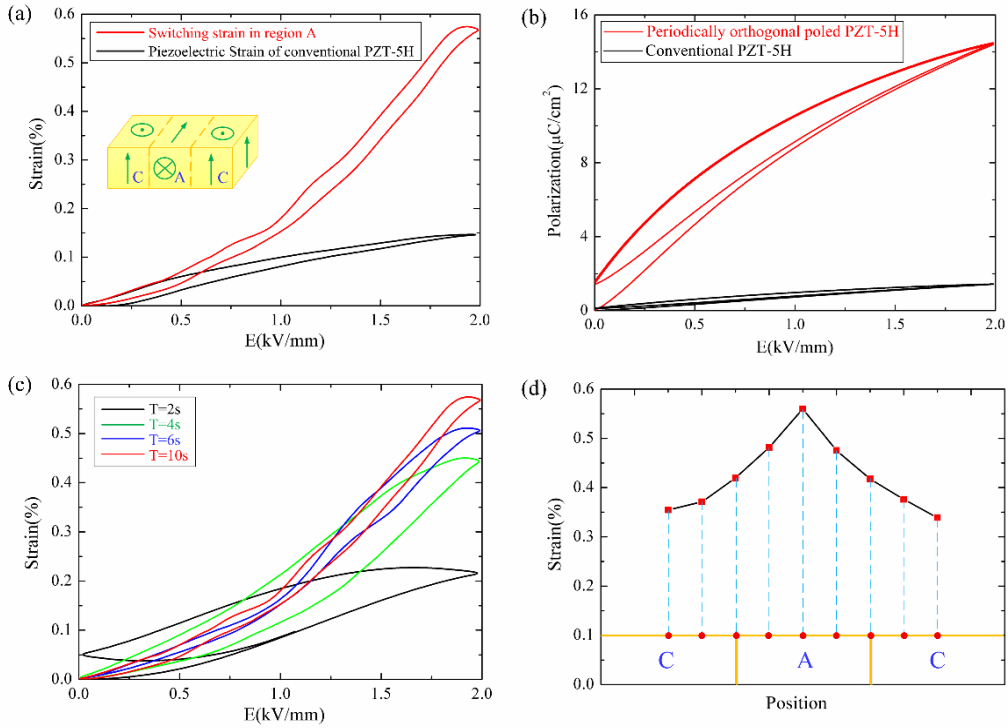

Fig.3 Actuation testing results on the periodically poled PZT-5H ceramics. (a) Electrostain curves in Region A and (b) polarization curves for the loading/unloading period of 10s (curves for the conventional PZT-5H is also plotted for comparison); (c) Electrostrains in Region A for different loading/unloading periods; (d) Maximum electrostrain profile along the length direction under 2kV/mm.

Fig.3 shows the actuation testing results on the periodically poled PZT ceramics under a uni-polar electric field up to 2kV/mm. Fig.3(a) and 3(b) are the electrostrain curves in the center of Region A and polarization curves for the loading/unloading period of 10s. For comparison, the electrostrain curve and polarization curve of a conventional PZT is also plotted. It can be seen from Fig.3(a) that the maximum switching strain in Region A can reach up to 0.57% at 2kV/mm, about 4 times of that in conventional PZT ceramics (which is about 0.14% at 2kV/mm). Furthermore, the large switching strain is totally reversible with small hysteresis. Fig.3(b) shows that the polarization variations in this periodically poled PZT-5H is very large compared to that of the conventional PZT-5H, which is mainly caused by the non-180° domain switching in

Region A.

Fig.3(c) shows the electrostrain curves in the center of Region A of the periodically poled PZT-5H with different loading/unloading periods (T). It can be seen that the electrostrain increases significantly with the increasing period. For T=2s, the maximum strain under 2kV/mm is only about 0.2% with large hysteresis. The maximum strain increases to 0.45% for T=4s, 0.51% for T=6s, and reaches 0.57% for T=10s. Meanwhile, the strain hysteresis decreases with the increasing period. Note that the eletostrain curves for T>10s almost overlapped with that for T=10s, indicating that T=10s is long enough to saturate the large reversible strain caused by domain switching. Due to the current limitation (2mA) of the high-voltage amplifier, testing with T<2s was not conducted.

Fig.3(d) shows the maximum electrostrain profile along the length direction under 2kV/mm with the loading/unloading period of 10s. It can be seen that the strain profile is almost symmetric about the center of Region A. The maximum strain under 2kV/mm decreases gradually from 0.57% at center of Region A to 0.32% at center of Region C, i.e., non-uniform strain was generated along the length direction. Note that the strain of 0.32% at center of Region C is also considerably larger than that of 0.14% for conventional PZT-5H.

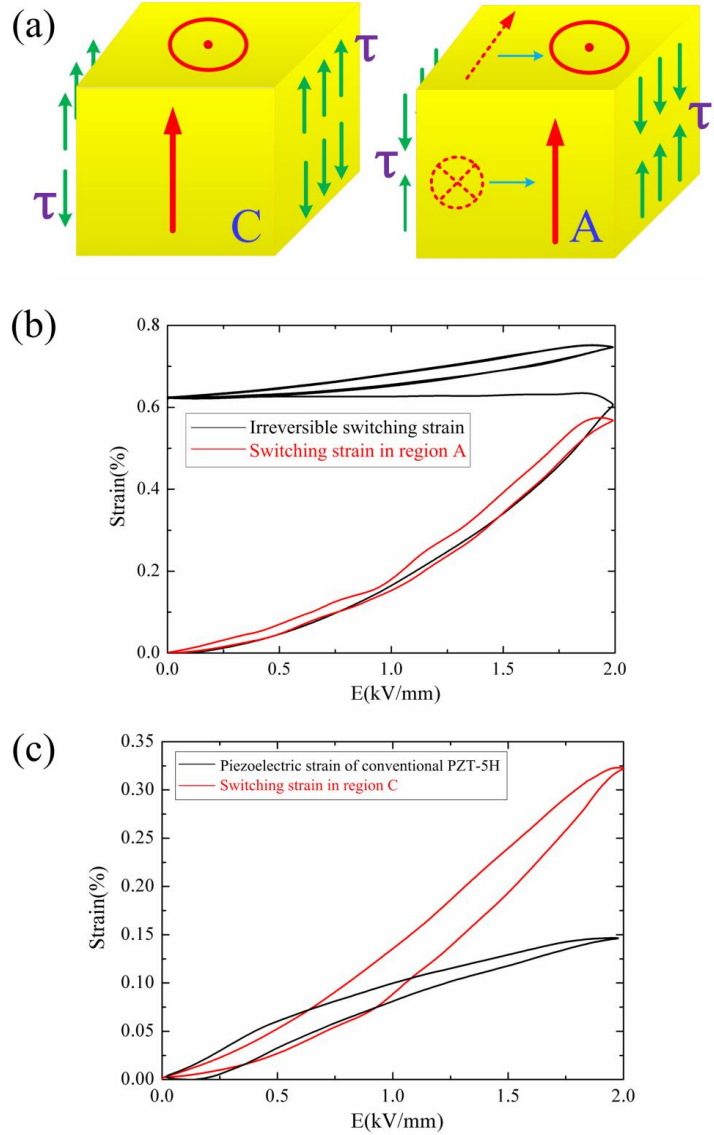

Fig.4 Mechanism of large reversible strain in the periodically orthogonal poled PZT-5H ceramics. (a) The generated shearing stress on the interface between Region C and Region A; (b) Electrostrain curves for the irreversible domain switching in conventional PZT-5H and for the constrained reversible switching strain in Region A; (c) Electrostrain curves for the piezoelectric strain in conventional PZT-5H and for the enhanced strain in Region C.

Now let us discuss the mechanism of large reversible switching strain in this periodically poled PZT-5H. When applying a large electric field on this PZT-5H sample, non-180° domain switching occurred, leading to large switching strain in Region A. While the piezoelectric strain in the adjacent Region C is quite small. Therefore, the misfit strain will generate shearing stresses on the interface between Region A and Region C, as shown in Fig.4(a). The interfacial shearing stress is apt to stretch Region C and compress Region A. Therefore, the constrained switching strain in Region A should be smaller than the free switching strain from the in-plane poled state to the vertical poled state (in Fig.1b), and the strain in Region C be larger than that of conventional PZT-5H. To verify this, the irreversible domain switching in shown in Fig.1(b) was also conducted in another PZT-5H sample and the electrostrain curves for the initial two cycles of uni-polar electric loading were plotted in Fig.4(b). For comparison, the constrained electrostrain curve (with T=10s) in the center of Region A is also presented in Fig.4(b). It can be seen that for the first cycle loading, these two curves almost overlapped, which may indicate that the interfacial stresses only has a slight influence on the non-180° domain switching at the center of Region A. Upon removing the electric field, the large interfacial stress will cause back domain switching in Region A, thus the large switching strain can be reversible. Since the interfacial stress is not constant but increases gradually with the applied electric field, both the domain switching and back switching are gradual processes, leading to very small hysteresis in the electrostrain curves, the underlying mechanism is similar to that of the reduced strain hysteresis in a relaxor/ferroelectric composite [19]. As to the irreversible domain switching PZT-5H sample, the irreversible strain even increase upon electric unloading and can reach about 0.63%. Thus, at the center of Region A, about 90% (=0.57/0.63) of the total switching strain can be reversible.

To understand the effect of interfacial stress on the electrostrain in Region C, the electostrain curve (T=10s) in the center of Region C was plotted in Fig.4(c) together with the electrostrain curve of a conventional PZT-5H. It can be seen that when the field is less than 900V/mm (slightly larger than the PZT's coercive field of 750V/mm), the two curves almost overlapped. By comparing the strain curves in both Fig.4(b) and 4(c) below 900V/mm, we can see that at this stage the misfit strain between Region C and Region A is very small (<0.1%), so the interfacial stress is also small and it has little effect on the piezoelectric strain in Region C. The misfit strain increases with the increasing electric field, so as to the interfacial stress, leading to the increased strain in Region C due to partial non-180°domain switching.

With regard to the loading frequency dependent switching strain as shown in Fig.3(c), we think it is due to the current limit (2mA) of the high-voltage amplifier. Because domain switching in this periodically poled PZT-5H is accompanied by large polarization variations (as shown in Fig.3b), in short-period loading, the amplifier cannot supply enough current thus charge into the sample, resulted incomplete domain switching. When the loading/unloading period is long enough (10s in this work), complete domain switching may accomplish and the switching strain saturates, i.e., further increasing the loading period will not enhance the switching strain any more. Quick domain switching is expected if using a large-current amplifier, which will be conducted in the future works.

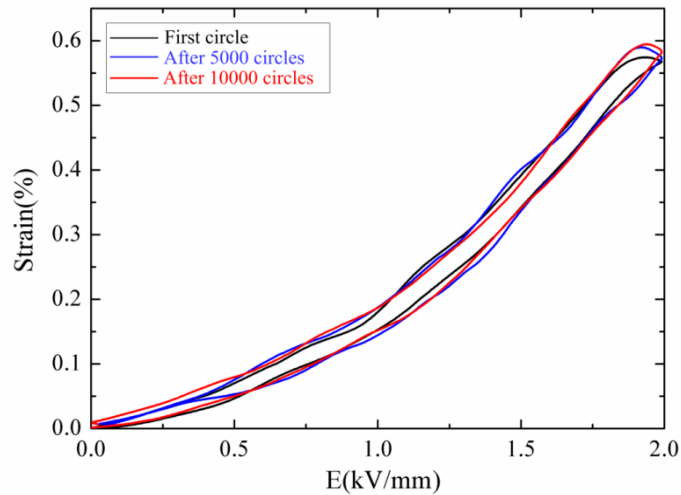

Fig.5 The electrostrain curves at center of Region A in the periodically poled PZT-5H after 1, 5000 and 10000 cycles of electric loading. (Loading/unloading period of 10s)

Fig.5 shows the electrostrain curves at center of Region A in the periodically poled PZT-5H after 1, 5000 and 10000 cycles of loading with the period of 10s. It can be seen that the large reversible strain keeps quite stable and even slightly increases (from 0.57% to 0.58%) after $10^4$ cycles of loading. The slightly increase in the maximum strain may be due to the interfacial stress relaxation after cycles of domain switching. The large, reversible and stable switching strain is very promising for next-generation large-strain actuators.

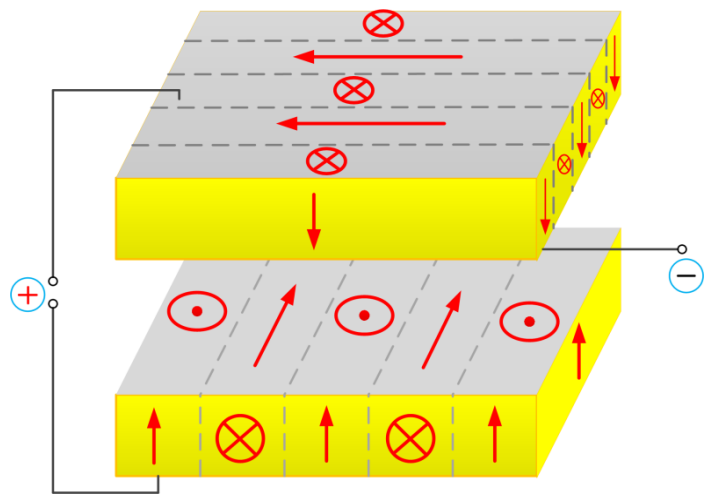

Fig.6 Schematic design of a multilayer actuator based on the periodically orthogonal poled ferroelectric ceramics. Arrow denotes poling directions.

Fig.6 shows the schematic design of a multilayer actuator based on the periodically orthogonal poled ferroelectric ceramics. The periodic directions of the neighboring layers are orthogonal, thus the maximum switching strain of each layer can always be employed. It is expected the layer thickness, the applied prestress, the width of periodic poling Region in each layer, etc., may considerably affect the performance of such a designed multilayer actuator, which will be systematically investigated in near future.

## 4. Conclusions

In summary, we tailored the domain structures in a lead titanate zirconate ceramic via periodically orthogonal poling and realized large actuation strain up to 0.57% via reversible domain switching. The interfacial shearing stresses were generated upon applying the electric field, which induced back domain switching during electric field unloading. The large actuation strain is very stable and even slightly increases after $10^4$ cycles of electric loading, which is very promising for next-generation large-strain actuators.

**References**

[1] K. Uchino, *Piezoelectric Actuators and Ultrasonic Motors* (Kluwer Academic, Boston, 1996)

[2] K. Otsuka, X. Ren, Prog. Mater. Sci. **50**, 511 (2005)

[3] M.Wun-Fogle, J.B. Restorff, A.E. Clark, and J. Snodgrass. IEEE Trans. Magn. **39**, 3408 (2003)

[4] S.-E. Park and T. R. Shrout, J. Appl. Phys. **82**,1804 (1997)

[5] Y. J. Wang, L.J. Chen, G.L. Yuan, H.S. Luo, J.F. Li and D. Viehland. Scientific Rep. **6**: 35120 (2016)

[6] E. A. McLaughlin, T. Q. Liu, and C. S. Lynch, Acta. Mater. **52**, 3849 (2004).

[7] K. G. Webber, R. Z. Zuo, and C. S. Lynch, Acta. Mater. **56**, 1219, (2008).

[8] D. Viehland and J. Powers, J. Appl. Phys. **89**, 1820 (2001)

[9] J. Rödel, W. Jo, K. T. P. Seifert, E.-M. Anton, T. Granzow, D. Damjanovic, J. Am. Ceram. Soc. **92**, 1153 (2009)

[10] W. Jo, R. Dittmer, M. Acosta, J. Zang, C. Groh, E. Sapper, K. Wang, J. Rödel, J. Electroceram. *29*, 71 (2012)

[11] X.M. Liu and X. Tan. Adv. Mater. **28**, 574 (2016)

[12] J. Fu, R. Zuo, H. Qi, C. Zhang, J. Li, L. Li, Appl. Phys. Lett. **105**, 242903 (2014)

[13] X. B. Ren, Nat. Mater. **3**, 91 (2004)

[14] Z. Y. Feng, O. K. Tan, W. G. Zhu, Y. M. Jia, and H. S. Luo, Appl. Phys. Lett. **92**, 142910 (2008)

[15] E. Burcsu, G. Ravichandran, and K. Bhattacharya, Appl. Phys. Lett. **77**, 1698 (2000).

[16] J. Shieh, J. H. Yeh, Y. C. Shu, and J. H. Yen, Appl. Phys. Lett. **91**, 062901(2007)

[17] Y.W. Li and F.X. Li. Appl. Phys. Lett. **102**, 152905 (2013)

[18] H.C. Miao and F.X. Li. Appl. Phys. Lett. **107**, 122902 (2015)

[19] C. Groh, D. J. Franzbach, W. Jo, K. G. Webber, J. Kling, L. A. Schmitt, H. Kleebe, S.J. Jeong, J.S. Lee, and J. Rödel, Adv. Funct. Mater. **24**, 356 (2014)